\documentclass{PoS}
\usepackage{epsfig}

\newcommand{\mypreprint}[1]{\begin{flushright} #1 \end{flushright}}
\newcommand{\be}{\begin{equation}}
\newcommand{\ee}{\end{equation}}
\newcommand{\ba}{\begin{eqnarray}}
\newcommand{\ea}{\end{eqnarray}}

\title{Decays of mesons with charm quarks on the lattice
\vspace{-3cm}\mypreprint{\small{\it DESY 07-168\\ IFUP-TH/2007-25}} 
\vspace{1.5cm}}

\ShortTitle{Decays of mesons with charm quarks on the lattice}

\author{
\speaker{A.~Ali~Khan}$^a$\thanks{Address after 
1/10/2007: Department of Physics, Faculty of Science, Taiz University, 
Taiz, Yemen Republic.}$\,\,\,$, V.~Braun$^a$,T.~Burch$^a$, 
M.~G\"ockeler$^a$, G.~Lacagnina$^b$, 
A.~Sch\"afer$^a$,   and G.~Schierholz$^{c}$ \\
$^a$  Institut f\"ur Theoretische Physik, 
 Universit\"at Regensburg, 93040 Regensburg, Germany \\
$^b$  Dipartimento di Fisica, 
  Universit\`a di Pisa and INFN, Pisa, Italy \\
$^c$  John von Neumann-Institut f\"ur Computing NIC, Deutsches 
  Elektronen-Synchrotron DESY,  15738 Zeuthen, Germany\\
  Deutsches Elektronen-Synchrotron DESY, 22603 Hamburg, Germany \\
        E-mail: \email{arifa.ali-khan@physik.uni-regensburg.de}}

\abstract{We investigate mesons containing charm quarks on fine lattices 
with $a^{-1} \sim 5$ GeV. The quenched approximation is employed using the Wilson 
gauge action at $\beta = 6.6$ and  nonperturbatively 
$O(a)$ improved Wilson quarks. We present results for decay constants 
using various interpolating fields and give preliminary results for 
form factors of semileptonic decays of $D_s$ mesons to light 
pseudoscalar mesons.}

\FullConference{The XXV International Symposium on Lattice Field Theory\\
		 July 30-4 August 2007\\
		 Regensburg, Germany}

\begin{document}

\section{Introduction}
There are  current experimental and theoretical activities to investigate
the decays of  heavy-light hadrons. Study of their weak decays is of interest 
for determining the CKM matrix of quark mixing, and there are theoretical and experimental 
efforts to test and overconstrain the Standard Model and find signatures of New Physics.

It is of interest to study
charmed hadrons on fine lattices where discretization effects
are very small. While unquenched data from very fine lattices is not yet
easy to obtain, the QCDSF collaboration has undertaken a quenched 
calculation of charmed and also bottom decay constants, weak matrix elements 
and meson spectra on a 
quenched lattice with an inverse lattice spacing of $a^{-1} \simeq 5$ GeV.
Results on heavy-light and light meson decay constants have been presented in a
previous paper~\cite{Ali Khan:2007tm}. A particularly interesting result
of this calculation was a relatively low value of the decay constant $f_{D_s}$. 
We have analyzed further matrix elements and would like to present
our findings here. 
In  Section~\ref{sec:simulation} we give a short description of simulation and analysis details.
Our calculation of the pseudoscalar decay constants using two different 
interpolating operators for the pseudoscalar meson is discussed  in 
Section~\ref{sec:decconstants}. 
In Section~\ref{sec:form} we give preliminary results for form factors of 
semileptonic decays of pseudoscalar heavy-light mesons to pseudoscalar 
light mesons. 
\section{Simulation parameters\label{sec:simulation}}
Our quenched lattices are generated using the Wilson gauge field action at 
$\beta = 6.6$. The lattice spacing, determined using the Sommer parameter
$r_0 = 0.5$ fm from~\cite{Necco:2001xg}, is $a = 0.040$ fm or 
$a^{-1} = 4.97$ GeV. The lattice size is  $40^3\times80$. For the
results reported here, we have analyzed 114 gauge field configurations.

We use $O(a)$ improved Wilson quarks. The value of the clover coefficient has 
been determined nonperturbatively by Ref.~\cite{Luscher:1996ug}. 
We work with three `light' and four `heavy' hopping parameters. 
Their values and the corresponding quark masses can be found in Table~\ref{tab:masses}.
\begin{table}[htb]
\begin{center}
\begin{tabular*}{0.90\textwidth}{@{\extracolsep{\fill}}lccccccc}
\hline
$\kappa $ & $0.13519$ & $0.13498$ & $0.13472$ & $0.13000$ & $0.12900$ & $0.12100$ & $0.11500$   \\
$a\tilde{m}_q$ & 0.0076 & 0.013 & 0.020 & 0.14 & 0.16  & 0.31 & 0.37 \\
$M$[GeV] & 0.55 & 0.69 & 0.87 & 2.8 & 3.2 & 5.6 & 7.2 \\
\hline
\end{tabular*}
\end{center}
\caption{Hopping parameters, the corresponding $O(a)$ improved quark masses 
(see Eq.~(\protect\ref{eq:aqm})) and approximate
pseudoscalar meson masses (denoted as $M$).
}
\label{tab:masses}
\end{table}

In our analysis we parameterize the light quark masses  using the  
$O(a)$ improved quark mass 
\be  
a\tilde{m}_q = (1 + b_m am_q), \label{eq:aqm}
\ee 
where $am_{q} = \frac 1 2 (\frac{1}{\kappa}-\frac{1}{\kappa_{crit}})$. 
The nonperturbatively determinated value of $b_m$ is taken from 
Ref.~\cite{Guagnelli:2000jw}. As a measure for the heavy quark masses
we use the mass of the heavy-light meson, where the light quark mass
is extrapolated to the average of the $u$ and $d$ quark mass. 

We calculate matrix elements of the $O(a)$ improved axial vector current
containing the quarks $q_1$ and $q_2$:
\be
A_4^I  =  Z_A(1+ b_A am_q) 
\left(A_4 + ac_A \partial_4 P\right),
\ee
where $A_4(x) = \overline{q}_1(x)\gamma_4\gamma_5q_2(x)$ is the local axial vector
lattice current,  and $P(x) = \overline{q}_1(x)\gamma_5q_2(x)$ the local
pseudoscalar density. With $\partial_4$ we denote the symmetric lattice 
derivative. We use the nonperturbatively determined values 
for $c_A$ from \cite{Luscher:1996ug} and for $Z_A$ from \cite{Luscher:1996jn}.
For the coefficient $b_A$ of the  
bare quark mass $am_q$, we use the value of Ref.~\cite{Sint:1997jx}, calculated in one-loop
perturbation theory, and the boosted lattice coupling. 

The $O(a)$ improved vector current is given by
\be
V_\mu^I = Z_V(1+am_qb_V)\left(V_\mu + 
i\,a\,c_V\partial_\nu T_{\mu\nu}\right) , \label{eq:vector}
\ee
where $V_\mu = \overline q_1 \gamma_\mu q_2$ and $T_{\mu\nu} = 
\overline q_1 \frac i 2 [\gamma_\mu,\gamma_\nu] q_2$.
We use for $Z_V$ and $b_V$ the nonperturbatively calculated results 
from~\cite{Bakeyev:2003ff}. For the correction due to the tensor current
we have only made a preliminary order-of-magnitude estimate, using a
rational interpolation~\cite{pleiter_thesis} of the nonperturbative values for
$c_V$ from~\cite{Guagnelli:1997db}.
\section{Pseudoscalar decay constants \label{sec:decconstants}}
We calculate the decay constant $f$ of pseudoscalar 
mesons $M$ at zero momentum from 
\be
f = \frac 1 M \langle 0| A_4^I | M \rangle.
\ee
Previously~(\cite{Ali Khan:2007tm}) we used operators of the type 
$P$ to project at the source onto the pseudoscalar meson.
\begin{table}[htb]
\begin{center}
\begin{tabular*}{0.95\textwidth}{@{\extracolsep{\fill}}|l|l|l|l|l|l|}
\hline
 \multicolumn{2}{|c|}{}
& \multicolumn{2}{c|}{$am_{PS}$} 
& \multicolumn{2}{c|}{$af^{(0)}$} \\
 \multicolumn{1}{|c|}{$\kappa_1 $}
& \multicolumn{1}{c|}{$\kappa_2 $}
& \multicolumn{1}{c|}{$A4P$}  
& \multicolumn{1}{c|}{$A4A4$}  
& \multicolumn{1}{c|}{$A4P$}  
& \multicolumn{1}{c|}{$A4A4$}  \\
\hline
0.11500 & 0.13519   & $0.8363(15)$  & $0.8350(14) $ & $ 0.0371(11)$ & $0.0367(15)$  \\
0.12100 & 0.13519   & $0.6676(13)$  & $0.6666(13) $ & $ 0.0417(14)$ & $0.0404(15)$  \\
0.12900 & 0.13519   & $0.4065(11)$  & $0.4058(12) $ & $ 0.0475(13)$ & $0.0465(14)$  \\
0.13000 & 0.13519   & $0.3685(12)$  & $0.3678(12) $ & $ 0.0478(13)$ & $0.0469(13)$  \\
0.11500 & 0.13498   & $0.8431(12)$  & $0.8415(12) $ & $ 0.0383(12)$ & $0.0377(14)$  \\
0.12100 & 0.13498   & $0.6747(11)$  & $0.6735(11) $ & $ 0.0429(12)$ & $0.0418(14)$  \\
0.12900 & 0.13498   & $0.4145(10)$  & $0.4137(09) $ & $ 0.0488(15)$ & $0.0480(14)$  \\
0.13000 & 0.13498   & $0.3765(09)$  & $0.3756(09) $ & $ 0.0490(13)$ & $0.0482(13)$  \\
0.11500 & 0.13472   & $0.8517(11)$  & $0.8503(11) $ & $ 0.0402(12)$ & $0.0393(14)$  \\
0.12100 & 0.13472   & $0.6836(10)$  & $0.6826(10) $ & $ 0.0446(13)$ & $0.0437(15)$  \\
0.12900 & 0.13472   & $0.4242(08)$  & $0.4234(08) $ & $ 0.0508(13)$ & $0.0497(14)$  \\
0.13000 & 0.13472   & $0.3866(08)$  & $0.3860(08) $ & $ 0.0507(13)$ & $0.0501(14)$  \\
\hline
\end{tabular*} 
\end{center}
\caption{\small Pseudoscalar heavy-light meson masses and decay constants at 
$O(a^0)$ from $\langle A_4P\rangle$ and $\langle A_4 A_4 \rangle$ 
correlation functions at the simulated hopping parameters.}
\label{tab:hdecay}
\end{table}
Since different correlation functions may be subject to different systematic  
errors, we
would like to determine the matrix elements of $A_4$ also using the 
temporal component of the axial vector current operator, i.e.\ from
correlation functions of the form
\ba
C_{A4A4}^{Si}(t) &=&  \sum_{\vec x} \langle A_4^i(\vec x, t) A_4^{S\dagger}(0) \rangle.
\label{eq:corre}
\ea
The index $i$ stands for local ($i=L$) or Jacobi 
smeared ($i=S$) operators.
If we write the decay constant as
\be
f = Z_A (1+b_A am_q)(f^{(0)} + ac_A f^{(1)}),
\ee
we find the unimproved contribution $f^{(0)}$ from the correlation function in
Eq.~(\ref{eq:corre}).

We extract masses and amplitudes from  single state fits to the correlation functions.  
The pseudoscalar meson masses and the results for $f^{(0)}$ are compared 
for heavy-light mesons at the simulated quark masses in 
Table~\ref{tab:hdecay} with the corresponding values 
from~\cite{Ali Khan:2007tm}.
The results agree within the statistical errors.
\begin{table}[htb]
\begin{center}
\begin{tabular*}{0.90\textwidth}{@{\extracolsep{\fill}}cc|cc}
\hline
\multicolumn{2}{c|}{From $A4PS$ correlators (Ref.\cite{Ali Khan:2007tm})} &
\multicolumn{2}{c}{From $A4A4$ correlators (this work)} \\
\hline
\multicolumn{4}{c}{Light meson decay constants} \\
$f_\pi$[MeV] & $f_K$[MeV] & $f_\pi$[MeV] & $f_K$[MeV] \\
140(4) & 152(4) &  138(6) & 150(5)  \\
\hline
\multicolumn{4}{c}{Heavy-strange meson decay constants} \\
$f_{D_s}$[MeV] & $f_{B_s}$[MeV] & 
$f_{D_s}$[MeV] & $f_{B_s}$[MeV] \\
$220(6)(5)(11)$ & $205(7)(26)(17)$ & $217(5)$ & $204(9)$  \\
\hline
\multicolumn{4}{c}{Heavy-light meson decay constants} \\
$f_{D}$[MeV] & $f_{B}$[MeV] & 
$f_{D}$[MeV] & $f_{B}$[MeV] \\
$206(6)(3)(22)$ & $190(8)(23)(25)$ & $202(6)$ & $191(9)$  \\
\hline
\end{tabular*}
\end{center}
\caption{Results on decay constants from different correlators. The first
error is statistical, the second error on the results 
from~\cite{Ali Khan:2007tm} is systematic and the third from the
 uncertainty in the experimental value of $r_0$.
We estimate the systematic errors in the new results
to be similar.}
\label{tab:decayconstants}
\end{table}

We fit the matrix elements to linear functions in the light quark masses and, 
where applicable, to quadratic functions in the heavy quark masses, using the 
method described in~\cite{Ali Khan:2007tm}. To determine the physical
light, strange and heavy quark masses we also use the same method as for the
central values given in~\cite{Ali Khan:2007tm}.
The decay constant results are listed in Table~\ref{tab:decayconstants}.  
We take over the errors for the heavy-light decay 
constants from~\cite{Ali Khan:2007tm}, including
uncertainties from setting the quark masses to the physical values, 
discretization effects, errors in  the nonperturbative renormalization
constants and from finite volume effects. A $10\%$ uncertainty
in the experimental value of $r_0$ is also included. 
We did not yet perform the same
reanalyses for the data from the $A_4A_4$ correlators
and just assume that the systematic errors are very similar.
We did not
attempt an estimate of the systematic error for the light meson decay 
constants. 

Our result for $f_{D_s}$ is still smaller than the recently updated
experimental value of the $D_s$ decay constant of $275(10)_{stat}(5)_{syst}$ 
MeV by CLEO~\cite{Stone:2007dy}.
\section{Semileptonic form factors \label{sec:form}}
We describe preliminary results for matrix elements of semileptonic decay 
of pseudoscalar heavy-light mesons ($`D'$) with mass $M_h$ and 
momentum $p_h$, to pseudoscalar light mesons ($`\pi'$) with mass $M_l$ 
and momentum  $p_l$. 
The decay goes via a vector 
current $V_\mu(x)$ with the matrix element
\be
{\cal M}_\mu = \langle \pi(p_l)|V_\mu(0)|D(p_h) \rangle.
\ee
We would like to determine the two form factors which are often used
to parameterize the matrix element
\be
{\cal M}_\mu = (p_h + p_l - \Delta q)_\mu F_+(q^2) + \Delta q_\mu F_0(q^2),
\ee
with $q = p_h-p_l$ and $\Delta = (M_h^2-M_l^2)/q^2$~\cite{Abada:2000ty}. 
\begin{figure}[thb]
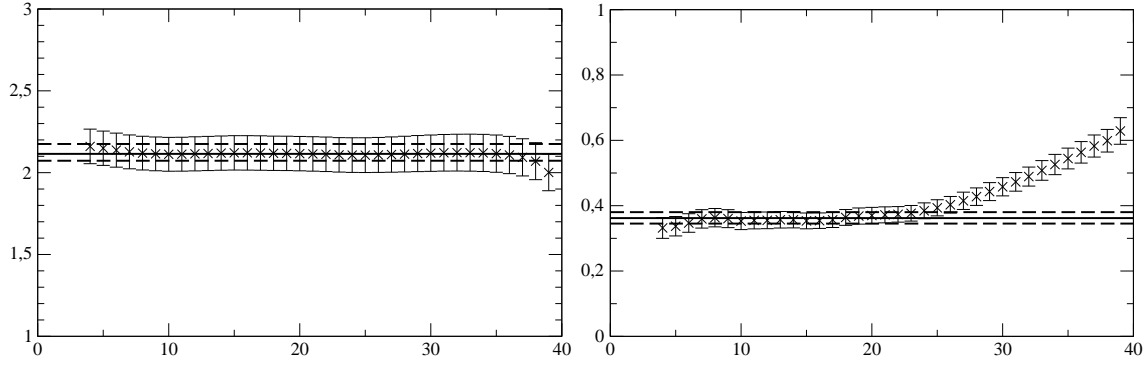

\begin{center}
\centerline{
\epsfig{file=threept4_pe1_p4.eps,width=7.5cm}
\epsfig{file=threeptk_pe1_p4.eps,width=7.5cm}
}
\vspace{-0.2cm}
\end{center}
\caption{`Divided' three point functions at the hopping parameter of the heavy
quark $\kappa_h = 0.12900$ and the hopping parameter $\kappa = 0.13498$ for the 
light and spectator quark. The meson momenta are  $\vec p_h = (1,0,0)$ and
$\vec q = (1,0,0)$ in units of $2\pi/L$. The crosses denote the data points,
the solid lines the central values and the dashed lines the error bars of the 
fit. 
 On the left we show $|{\cal M}_4|$ 
and on the right $|{\cal M}_1|$.
}
\label{fig:effmass}
\end{figure}
The matrix element can be extracted from the three point correlation function
of a pseudoscalar light meson at time zero, a vector current at time $t_x$
and a pseudoscalar heavy-light meson at time $t_y = T/2$, where $T$ is the 
time extent of the lattice:
\be
C_\mu^{(3)}(t_x,t_y) = \sum_{\vec{x},\vec{y}}
e^{-i\vec{p}_h\cdot \vec{y}}e^{i\vec{q}\cdot \vec{x}}
\langle P^S_h(\vec{y},t_y)V_\mu(\vec{x},t_x)P^S_l(0) \rangle.
\ee
$P_h^S$ and $P_l^S$ are Jacobi smeared operators with the structure 
$\overline{q}\gamma_5q_{l}$ and $\overline{q}_{h}\gamma_5q$. With
$q_h$ we denote the heavy quark, with $q_l$ the light quark, and with $q$ the 
spectator quark. The vector current is of the type
$V_\mu = \overline q_l\gamma_\mu q_h$.

The large time behavior $(0 \ll t_x \ll t_y )$ or $(0 \ll t_y \ll t_x )$ 
of the three point function is given by
\ba
C_\mu^{(3)}(t_x,t_y) & \longrightarrow & 
\frac{Z^S_l}{2E_l}
\frac{Z^S_h}{2E_h}e^{-E_l t_x}
e^{-E_h (t_y-t_x)}\;\langle \pi(p_l)|V_\mu|D(p_h)\rangle\,,\; 
t_x < t_y\,, \nonumber \\
C_\mu^{(3)}(t_x,t_y) &\longrightarrow & \pm
\frac{Z^S_h }{2E_h}\frac{Z^S_l}{2E_l}
e^{-E_l (T-t_x)}e^{-E_h (t_x-t_y)}\;
\langle \pi(p_l)|V_\mu|D(p_h)\rangle\,,
\; t_x > t_y\,, \nonumber 
\ea
where the prefactors are given by
$Z^S_h = |\langle 0|P^S_h|D(\vec{p}_h)\rangle|$ and
$Z^S_l = |\langle 0|P^S_l|\pi(\vec{p}_l)\rangle|$. The correlation functions
are symmetric or antisymmetric around $T/2$. We bin over the contributions
from $t_x > t_y $ and  $t_x < t_y $, and divide the three point functions by the 
prefactors extracted from fits to smeared-smeared two point functions in
a bootstrap loop. 
The `divided' three point function is fitted to a constant to extract the 
matrix element. An example is shown in Fig.~\ref{fig:effmass}.
In this preliminary analysis we considered in particular momentum 
combinations where the $D$ and the $\pi$ both 
have spatial momenta $\leq 2\pi/L$,  where $L$ is the spatial lattice extent,
and the momenta are aligned, i.e.:
\be
\begin{array}{cccccl}
\vec{p}_h &=& (0,0,0) \,, &  -\vec{q} &= & (0,0,0)\,,\;  (1,0,0)\,,\;  (0,1,0)\,,\; 
              (0,0,1)\,,\;   (-1,0,0)\,,       \\
\vec{p}_h &=& (1,0,0) \,, &  -\vec{q} &= & (0,0,0)\,,\;  (-1,0,0)\,,\;  
(-2,0,0)\,, \\
\end{array}
\ee
in units of $2\pi/L$.
The kinematic conditions are such that for each heavy
and light meson mass, each different combination of $\vec{p}_h$
and $\vec q$ leads to a different value of 
$q^2 = (E_h-E_l)^2 - \vec{q}^2$. To extract the form factors
for each given value of $q^2$ separately, without making a guess for the $q^2$
dependence of the form factors, we only have the
four Lorentz components of the matrix elements ${\cal M}_\mu$ at our
disposition. 

We have exploited the fact that the form factors appear in the
spatial and temporal components of ${\cal M}_\mu$ with different momentum 
prefactors, so we obtain a linear equation system which we can solve for 
$F_0$ and $F_+$.

\begin{figure}[thb]
\begin{center}
\epsfig{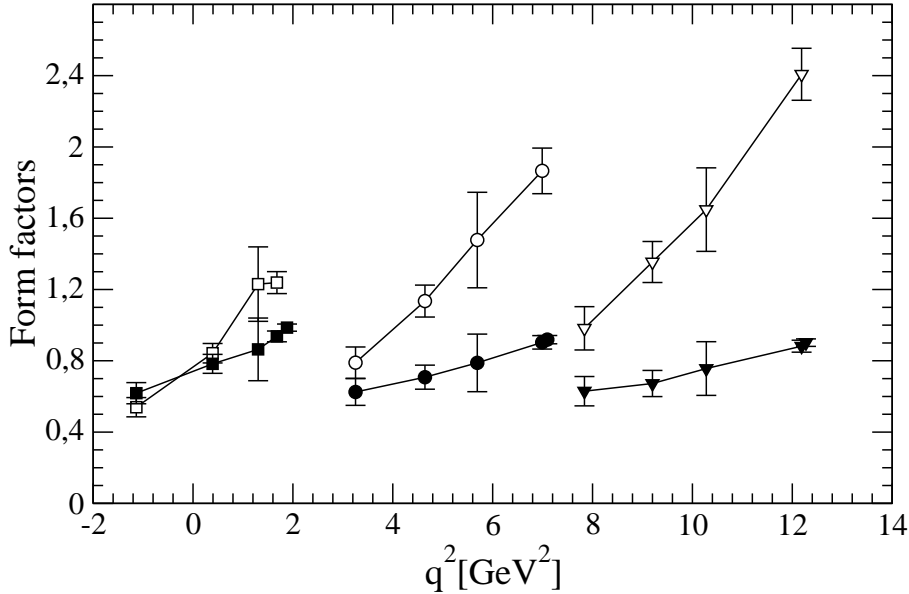}
\vspace{-0.2cm}
\end{center}
\caption{Form factors for semileptonic decays for various heavy quark masses.
The light and spectator quark 
hopping parameters are fixed to the $\kappa$ value $0.13498$.
Filled symbols denote $F_0$, open symbols $F_+$. Squares correspond to 
$\kappa_h = 0.129$ ($M_h = 2.1$ GeV), circles to $\kappa_h = 0.121$ ($M_h = 
3.4$ GeV), and triangles to $\kappa_h = 0.115$ ($M_h = 4.2$ GeV).
}
\label{fig:form}
\end{figure}

Results for the form factors at several heavy $\kappa$ values and with the
light and spectator quark mass close to the strange quark mass are shown 
in Fig.~\ref{fig:form}. For a preliminary estimate of the statistical error 
we use  error propagation of the bootstrap 
errors of the spatial and temporal components of
${\cal M}_\mu$. In the results shown in the Figure, the $O(a)$ correction to the
vector current (see Eq.~(\ref{eq:vector})) is not yet included. We
have however estimated its magnitude to be at most a few
percent for heavy quark masses around the 
charm. Although the quark masses are not tuned to the physical values matching
$K$ mesons, and the results correspond to decays of $D_s$ to
pseudoscalar strange mesons, we attempt to compare them to experiment and 
other lattice results. For $\kappa = 0.12900$, close to $\kappa_{charm}$, we find
 $F_0(0) \approx F_+(0) \approx 0.75 $. For $D \rightarrow K$ decays, a recent experimental
determination from  BABAR has found a value of 
$0.727(11)$~\cite{Aubert:2007wg}, a 
lattice calculation with three flavors of dynamical 
quarks on a coarser lattice using Fermilab heavy quarks
quotes $F_+(0) = 0.73(8)$~\cite{Aubin:2004ej}, and a quenched calculation on a coarser 
lattice quotes $F_+(0) = 0.66(4)$~\cite{Abada:2000ty}. So there seems to be
at least a rough agreement, but for a more precise result the tuning of quark
masses to the physical values will have to be performed.
\section*{Acknowledgements}
The numerical calculations have been performed on the Hitachi SR8000 at LRZ
Munich. This work was supported by DFG (Forschergruppe
Gitter-Hadronen-Ph\"anomenolo\-gie) and GSI. 
A.A. thanks the DFG and ``Berliner Programm zur F\"orderung der
Chancengleichheit f\"ur Frauen in Forschung und Lehre''
for financial support.

\end{document}